\documentclass[twocolumn,showpacs,preprintnumbers,amsmath,amssymb]{revtex4}

\usepackage{graphicx}
\usepackage{dcolumn}
\usepackage{bm}

\begin{document}

\preprint{APS/123-QED}

\title{Advanced Detection of Information in Optical Pulses with Negative Group Velocity}

\author{Ulrich Vogl$^{\ast}$}

\author{Ryan T. Glasser}
\author{Paul D. Lett}
\affiliation{
{Quantum Measurement Division, National Institute of Standards and Technology and}\\
Joint Quantum Institute, NIST \& the University of Maryland, Gaithersburg, MD 20899 USA\\
}

\date{\today}

\begin{abstract}

In this letter we experimentally demonstrate that the signal velocity, defined as the earliest time when a signal is detected above the realistic noise floor, may be altered by a region of anomalous dispersion.  We encode information in the spatial degree of freedom of an optical pulse so that the imprinted information is not limited by the frequency bandwidth of the region of anomalous dispersion. We then show that the combination of superluminal pulse propagation and realistic detectors with non-ideal quantum efficiency leads to a speed-up of the earliest experimentally obtainable arrival time of the transmitted signal even with the overall pulse experiencing unity gain.  This speed-up is reliant upon non-ideal detectors and losses, as perfect detection efficiency would result in the speed of information being equal to the speed of light in vacuum, regardless of the group velocity of the optical pulses.     
\end{abstract}

\pacs{42.25.Bs,42.50.Gy,89.70.-a,42.65.Hw}

\maketitle

The possibility of optical pulse-propagation with a group velocity ($v_g$) greater than the vacuum  speed of light in materials having anomalous dispersion  has drawn continued interest since its early consideration \cite{1905Laue,1907Sommerfeld,1914Sommerfeld, 1914Brillouin}.
 Theoretical and experimental treatments have both shown that the speed of the peak of a pulse, which is associated with the group velocity, can be altered in a dispersive medium. Virtually any value of $v_g$ can be realized,  spanning from stopped light to negative group velocities. The speed of information that can be conveyed with a pulse is, however, associated with its front velocity $v_f$, i.e. the speed of the boundary separating the region in which the field vanishes identically from the region in which the field assumes nonzero values or, more generally, the speed of points of non-analyticity in the temporal waveform, and on a fundamental level $v_f$ cannot be altered by a dispersive medium \cite{1914Brillouin,garrett1970,chiao1994,2002Wynne,neifeld2003,boyd2003,zhang2011}. 

Research on anomalous group velocities in dispersive media has always been closely concerned with the associated speed of information transfer \cite{2000Peatross,2002Wynne,neifeld2003, boyd2003,  zhang2011}.
 Previous experiments investigating this question typically encode a bit of information in the time domain by superimposing a non-analytic step-function on a Gaussian pulse \cite{1998Mitchell,neifeld2003}.  The speed at which this point of non-analyticity propagates and is detected was shown to be $\leq c$, where $c$ is the speed of light in vacuum.  
 In these experiments the bandwidth of anomalous dispersion used is finite and relatively small ($\sim$\,MHz).  By introducing a sharp feature into the temporal pulse envelope, higher frequency components in the Fourier transform of the pulse become non-negligible and are outside of the bandwidth of anomalous dispersion and do not propagate superluminally.  

Here, in a complementary experiment, we use temporally smooth pulses and encode information in the spatial domain.  This allows us to investigate the arrival time of the spatial information while the majority of the pulse Fourier transform fits comfortably inside the anomalous dispersion regime. 
Also, while the faithful detection of a signal in the time domain requires a minimum sampling time depending on the temporal bandwidth of the signal, the detection of spatial information can be achieved in parallel and is limited by the detector specifications \cite{Shannon}.

In every practical information channel, noise and the detection
efficiency must be considered when discussing signal detection.
It has been shown \cite{ChiaoNoise,Molotkov2010} that, given ideal detection efficiency,  
the speed of information transfer through superluminal media is limited to $c$ due to the added (quantum-)noise, which usually leads to an increased detection latency.
For a non-ideal detection
efficiency, it has been suggested that an effective advancement of the earliest
possible detection of a signal may be achieved with superluminal
media \cite{exhibitA}. In this paper, we experimentally demonstrate that, given a realistic detector with sub-unity detection efficiency,
the detection of spatial information can be
advanced by using a superluminal channel, even when the total gain of the medium is one. 
While the advancement does not constitute the arrival of information travelling faster than c, it does constitute a movement of the detection closer to the information front.
In addition, any noise
added to the signal during superluminal propagation necessarily
limits the speed of information transfer as well.

\begin{figure*}
\includegraphics[width=18cm]{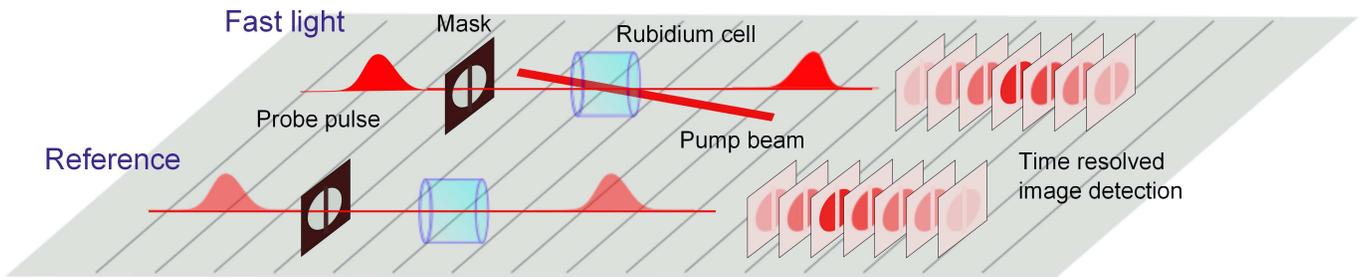}
\caption{\label{fig:epsart} (Color online) Experimental scheme for the arrival time measurement of spatial information in optical pulses. Depicted are the two situations for acquiring the reference pulse, and when the fast light medium is switched on. Probe pulses are shaped with a mask and injected into the rubidium cell at an angle of $\approx1^{\circ}$ relative to the pump beam.  A fast gated camera is used to detect and temporally resolve the pulses. The gating time of the camera intensifier can be lowered to 2.44\,ns and the gate delay is then swept temporally to
acquire images during the arrival of the pulse.  Images of the reference pulse are taken without the pump beam present. 
}
\end{figure*}

In our experiment we generate the dispersion with the steep negative slope required to produce fast light using a four-wave mixing scheme (4WM) in hot $^{85}$Rb vapor.
A strong pump beam (200\,mW) is sent through a vapor cell and a weak probe pulse, detuned by +3\,GHz, is injected at a small angle. The 4WM gain amplifies the probe pulse and generates a conjugate pulse at -3\,GHz from the pump in a separate spatial mode.
For these experiments we only study the probe pulse.
The resulting regions of strong anomalous dispersion form on the wings of the gain lines  \cite{2012Glasser,2012us}. 
 A detailed description of the superluminal pulse generation can be found in \cite{2012Glasser}.
A sketch of the experimental setup is shown in Fig.\,1.
   Temporally Gaussian pulses  (full width at half maximum (FWHM) of $200$\,ns) are created with an arbitrary waveform generator controlling an acousto-optical modulator and used as the probe.
As shown in Fig\,1, a mask is inserted in the collimated probe beam path in order to imprint spatial information on it.
   After the cell the amplified probe beam is singled out from the other light fields involved in the 4WM process and monitored by a fast gated ICCD (intensified charge-coupled device) camera. The intensifier of the camera is synchronized with the arbitrary waveform generator and is internally gated with a gate width of 2.44\,ns. The delay between the synchronizing signal and the gated time window can be swept to acquire the full temporal profile of the incoming pulse,
  allowing us to take temporal slices of the pulses.

   In this work we concentrate on the region of anomalous dispersion associated with the gain line for the injected beam, for which we measure a negative group index $n_g=\frac{c}{v_g}$ of up to -2400, where the group index is connected to the derivative with respect to the frequency $\nu$ of the refractive index by
 $n_g=n(\nu)+\nu dn(\nu)/d\nu$.
We use the pulse peak for the determination of $v_g$ (group-dispersion due to the higher order terms in the group index is clearly visible in the distortion of the pulses, leading to a larger advancement of the trailing edge of the pulse than the leading edge).
In Fig.\, 2b we show the arrival time of a spatially-structured probe pulse, integrated over its spatial profile, with the pump off (dashed blue line) and under fast light conditions (solid red line).
 The observed temporal reshaping of the superluminal pulse can be predicted 
 by applying the transfer function of the anomalous dispersion medium to the applied pulse form \cite{2003Mackereshaping}. 
Under fast light conditions the pulse is determined to have a negative group velocity.  This can be understood by considering its relation to a pulse's arrival time delay $\Delta T=\frac{L}{v_{g}}-\frac{L}{c}$ after propagating a distance $L$.  When the time delay and group velocity are negative, a pulse can appear to exit the medium at a time $T$ sooner than if it had traversed the same distance in vacuum.   We obtain a maximum pulse advancement of 90\,ns, which, with a cell length of 1.7\,cm, corresponds to a group velocity of $v_{g}=-\frac{1}{2400}c$, and a relative pulse peak advancement of $\approx 47\,\%$ of the original pulse width.

We now investigate how the information that is contained in an image is affected when it experiences a region of anomalous dispersion and when the carrier pulse propagates with a superluminal group velocity.
For this purpose we mask the pulse spatially with a single black stripe in the Gaussian probe beam cross section, as shown in Figs.\,2a,b.
Spatial integration over the full pulse cross section shows a pulse peak advancement of 90\,ns.  The advanced image shows inhomogeneous gain along the horizontal axis which we attribute  to the imperfect pump beam shape and angular dispersion that affects the phase-matching of the 4WM.
For the evaluation of the stripe visibility we choose a region of the image where the total integrated pulse intensity is equal for both the reference and the superluminal pulse, which allows us to set aside effects that result from amplification of the pulse, since the total gain is 1.

\begin{figure}
\includegraphics[width=7.2cm]{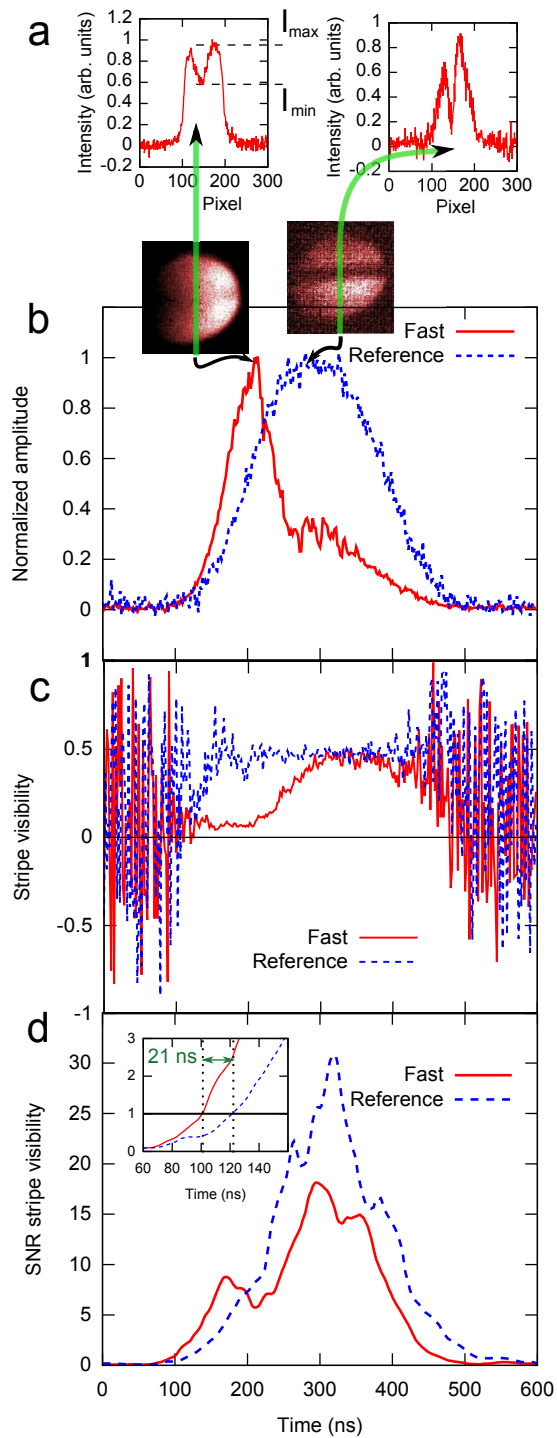}
\caption{\label{fig:epsart}
 (Color online) (a) A single dark stripe on a bright background is used as a test mask. Integration over the full pulse cross section shows a pulse peak advancement of 90\,ns. (b) Arrival time for the spatially integrated intensity of the pulses for the reference pulse without 4WM (dashed blue line) and the superluminal pulse (solid red line).
 (c) The stripe visibility $M=\frac{I_{\text{max}}-I_{\text{min}}}{I_{\text{max}}+I_{\text{min}}}$ of the transmitted pulses. A cross-section is taken through the image along the  vertical axis to analyse the double-peak structure in the spatial mode. (d) The SNR for the stripe visibility defined in the text for the advanced and the reference pulse as a function of time. In the inset we show the leading edge of the pulses where the SNR crosses the threshold of $1$.
 }
\end{figure}

We take a cross section through the image along the vertical axis to analyze the double-peak structure in the spatial mode as indicated in Fig.\,2a. We extract the average intensity of the region of the fringe maxima $I_{\text{max}}$ and the relative minimum $I_{\text{min}}$ in the center, and choose the stripe visibility  $M=\frac{I_{\text{max}}-I_{\text{min}}}{I_{\text{max}}+I_{\text{min}}}$ as the figure of merit. To determine $I_{\text{max}}$ and $I_{\text{min}}$ we calculate the spatially averaged intensity over an area of 3 x 90 pixels for each frame. 
The time-dependent stripe visibility that is obtained is shown in Fig.\,2c.
During the time when there is no pulse we see
strong fluctuations, since only dark noise is present.
At a gate delay of $\approx$120\,ns the noise drops drastically and the stripe becomes visible. The stripe visibility for the superluminal pulse is, for almost all times, lower than the visibility of the reference pulse, which can be attributed to the added noise from various processes, including gain. In Fig.\,2d we show the temporal evolution of the signal-to-noise ratio (SNR) of the visibility for the advanced and the reference beam, which we define as $\text{SNR}(t)={M(t)^2/\Delta M(t)^2}$, where $\Delta M^2(t)$ is the measured standard deviation of the stripe visibility. We evaluate $\Delta M^2(t)$ as the running standard deviation of the ten previous frames. The threshold of confidence for the detection of the spatial information in the pulse, i.e., when the signal exceeds the noise, is reached for the fast pulse 21$\pm$3\,ns earlier than for the reference pulse (uncertainties are combined systematic and statistical uncertainties, determined primarily by the gate width of the camera). Compared to the initial pulse width of 190\,ns, this is an 11\% relative advancement of the detection time compared to the reference pulse propagating with group velocity $c$.

We now want to compare the above results with a model that accounts for the added noise during the pulse propagation, which is especially important for the interval when the signal leaves the noise floor \cite{ChiaoNoise,Molotkov2010,exhibitA,2003Centini}.
To account for the added noise that the light pulse picks up while passing through the rubidium cell, we treat the system as a phase-insensitive amplifier, similar to the treatment in \cite{Milonni2007}:
$\hat{a}_{out}=g\hat{a}_{in}+\sqrt{|g|^2-1} \hat{b}^{\dagger}$.
Here $\hat{a}_{in}$ and $\hat{a}_{out}$ denote the input and the output of the amplifier, $\hat{b}=g\hat{a}+\hat{L}^{\dagger}$ is the cumulative noise source operator where $\hat{L}^{\dagger}$ denotes the excess noise and $|g|^2$ the power gain factor.
The photon number operator for the input is $\hat{n}_{in}=\hat{a}^{\dagger}\hat{a}$ and for the output $\hat{n}_{out}=|g|^2\hat{a}^{\dagger}\hat{a}+g^*\hat{a}^{\dagger}\hat{L}
^{\dagger}+g\hat{a}\hat{L}+\hat{L}^{\dagger}\hat{L}$.
This yields the expectation value $\langle \hat{n}_{out}\rangle=G \langle \hat{n}_{out} \rangle +G -1$, where $G=|g|^2$.
We have direct experimental access to $G$ via $G(t)=\frac{I_{out}(t)}{I_{in}(t)}$, as the gain line that is used is not accompanied by absorption.
The intensity fluctuations of the output are given by
$\langle \Delta \hat{n}^2_{out}\rangle=G^2 \langle \Delta \hat{n}^2_{in}\rangle+G(G-1)(\langle \Delta \hat{n}_{in}\rangle + 1)$.

To obtain a model for the stripe visibility $M$ we have to take into account 
the different photon numbers at the spatial intensity maxima and minimum:
\begin{equation}
M=\frac{\langle \hat{n}_{max}\rangle  - \langle \hat{n}_{min}\rangle  -D }
{\langle \hat{n}_{max}\rangle +  \langle \hat{n}_{min}\rangle -D},
\end{equation}
where $D$ accounts for the detector threshold.

We characterize the arrival time by using the information obtained in earlier frames to gain statistics for increasing time, following \cite{ChiaoNoise}. We integrate the signal-to-noise ratio starting from time zero, $\int \text{SNR}(t) dt$, where we take into account the different background noise levels and subtract them from the raw data before integrating.
The result is shown in Fig\,3a. The interesting part is
the region where $\text{SNR}_{\text{fast}}>\text{SNR}_{\text{reference}}$ at $t\approx$ 100\,ns to 200\,ns, showing
that, for the given experimental setup, we can detect the stripe signal earlier for a superluminal pulse.
 The advanced detection comes with the trade off that the total stripe visibility integrated over the whole pulse length decreases due to added noise from the 4WM process.

\begin{figure}
\includegraphics[width=6.8cm]{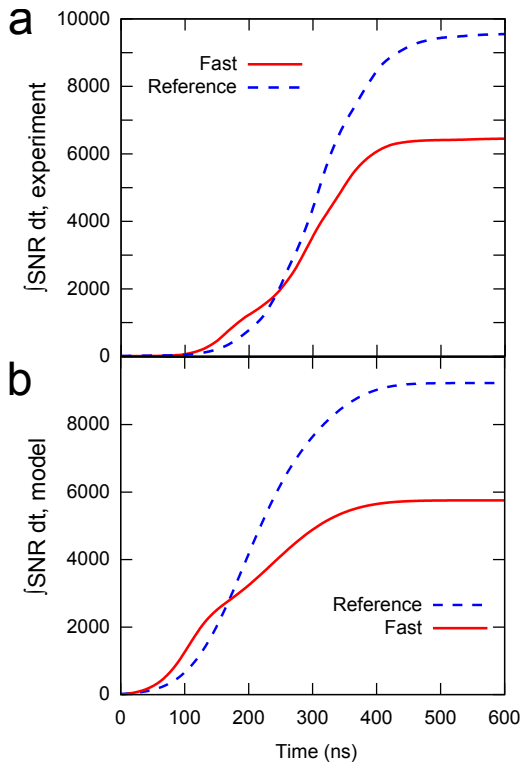}
\caption{\label{fig:epsart}
(Color online) Integrated stripe visibility signal. Panel (a) shows the integration $\int\text{SNR}(t) dt$ performed for the experimental data shown in Fig.\,3d. A model based on a phase-insensitive amplifier is shown for comparison in panel (b).}
\end{figure}

In Fig.\,3b we show a model for $\int\text{SNR}(t) dt$ based on Eq.\,1.
For a Gaussian pulse in time, both $I_{max}$ and $I_{min}$ of the imprinted stripe in the image would also be Gaussian in time with the half width $\tau$: $I(t)=I_0(x,y)*exp(-t^2/\tau^2)$.
We assume that the input pulse is shot noise limited and that the detector noise $N_i$ of the camera can be described pixel-wise: $I_i(x,y)=I_{i0}(x,y)+N_i(x,y)$, where $i$ denotes the pixel number.
For a Gaussian pulse, the time-integrated signal reads as $\int I dt\sim \textit{Erf}(t/\tau)$.

Our experimental reference pulse is well fit by a Gaussian, while the superluminal pulse after travelling through the region of anomalous dispersion can be well described by a compressed Gaussian pulse.
The model shown in Fig.\,3b uses pulse forms fitted to the data as in Fig.\,2b, i.e., an initial pulse width of 190\,ns, a total photon number per pulse of $\approx 3.8 \cdot 10^6$, and an overall detection efficiency of 30\,\%. 
The model reproduces the experimental observations reasonably well, as it predicts for a non-perfect detector a time window near the beginning of the pulse where the detection by this criteria would be earlier for the pulse that has transited the fast-light medium.
 It also predicts, that for  a perfect detector, no advanced detection is possible.
Depending on the detection threshold and the advancement of the pulse, we can find a time window when the spatial signal 
encoded on the advanced pulse can be detected at a given SNR sooner than for a similar signal on the reference pulse.
The pulse advancement, as demonstrated here, allows a signal to rise above a given SNR detection threshold earlier than in the case of a pulse travelling with group velocity of $c$. 
We have demonstrated this earlier signal detection using a nearly-Gaussian pulse shape, and under conditions of unity gain within the pulse bandwidth and a gain not exceeding ten outside of the region of the pulse spectrum.
This scheme is of relevance for signal transmission where a gain greater than one would saturate the detector.
  While we performed this experiment with a single stripe pattern representing the information, the implementation of a larger alphabet and denser spatial coding is possible and is mainly limited by the spot size of the beam diameter and the pump power.

In this work we have used superluminal spatially multi-mode images with a large negative group velocity to investigate the arrival time of optical signals.  This approach is less sensitive to pulse reshaping and distortion than a signal encoded in the time domain. We show that for a given realistic detector efficiency a defined spatial signal can be detected earlier if the signal beam is passed through a fast light medium than if it traverses the same distance in a medium without anomalous dispersion (for the case of unity gain).
The scheme used here of encoding information in a degree of freedom that is not affected by the limited temporal frequency bandwidth of the anomalous dispersion allows a conceptually clearer approach to determine the velocity of information in optical pulses than previous attempts \cite{2003Buettiker}.\\

\begin{acknowledgments}
We thank W.D. Phillips for discussions.
This work was supported by the Air Force Office of
Scientific Research. Ulrich Vogl would like to
thank the Alexander von Humboldt Foundation. This research was performed while
Ryan Glasser held a National Research Council Research
Associateship Award at NIST.
\end{acknowledgments}



\end{document}